\documentclass[twocolumn,showpacs,preprintnumbers,amsmath,amssymb]{revtex4}
\usepackage{graphicx}
\usepackage{dcolumn}
\usepackage{bm}
\begin{document}
\newcommand{\etal}{{\it et al. }}
\newcommand{\ie}{{\it i. e. }}
%
\title{Parity Effect and Tunnel Magnetoresistance of
  Ferromagnet / Superconductor / Ferromagnet Single-Electron Tunneling
  Transistors}

\author{Hiroshi Imamura, Yasuhiro Utsumi and Hiromichi Ebisawa}
\affiliation{Graduate School of Information Sciences, Tohoku University,
  Sendai 980-8579, Japan}
\pacs{72.25.-b,73.23.Hk,74.50.+r,73.23.-b}
%
%
\begin{abstract}
  We theoretically study the tunnel magnetoresistance(TMR) of
  ferromagnet / superconductor / ferromagnet single-electron
  tunneling transistors with a special attention to the parity effect.  
  It is shown that in the plateau region, there is no spin accumulation
  in the island even at finite bias voltage.    However, the
  information of the injected spin is carried by the excess electron
  and thus the TMR exists.  The spin relaxation rate of the excess
  electron can be estimated from the TMR.  We also show that the
  TMR increases with decreasing the size of the superconducting island.
\end{abstract} 

\maketitle
%
Single-electron tunneling(SET) transistor is a key device of ``single
electronics'' since the transfer of a single electron can be
controlled by gate and bias voltages \cite{sct_book,sohn_book}.  
The important quantity of the
SET transistor is the electrostatic energy of excess electrons in the
island called ``charging energy'', which has a significant effect on
charge transport, \ie, the Coulomb blockade(CB)~ \cite{sct_book}.  In
the CB region, sequential tunneling, where tunneling events in each
junction occur independently, is blocked at $T=0$ due to the increase
of the charging energy.  
Recently much attention has been devoted to the SET transistor
with a superconducting island and normal conducting electrodes(N/S/N)
\cite{fazio_book,ralph_book,averin1992,tuominen1992,eiles1993,utsumi2002}.
In a
superconducting island, where Cooper pairs form the condensate, the
addition of one extra electron costs the superconducting gap energy
$\Delta$.
This leads to the parity effect: physical properties of the system
depend on the parity of the electron number in the island.
In SET transistors consisting of normal conducting islands and
electrodes(N/N/N) the current depends $e$ periodically on the gate
charge.  However, the tunneling current of the N/S/N SET transistor is
expected to be 2$e$ periodic in the gate charge 
below the crossover temperature
due to the parity effect\cite{fazio_book}.
The clear signature of 2$e$ periodicity for the N/S/N SET transistor
was observed\cite{tuominen1992,eiles1993}.

The research on the N/S/N SET transistor has focused primary on the
charge degrees of freedom of electrons, by contrast, its spin degrees of
freedom have not yet received much attention.  However, an increasing
number of researches on spin-electronics show that the spin of
electron offers unique possibilities for finding novel mechanisms for
future spin-electronic devices \cite{maekawa_book,spinics_book}.  
The spin-polarized current injected from a ferromagnetic(F) electrode
into the N or S island gives rise to a nonequilibrium spin density in
the island.  In the F/N/F SET transistor, the tunnel magnetoresistance(TMR)
is brought about by spin accumulation in the
island \cite{imamura1999,brataas1999,barnas1999,korotkov1999}.
Recently Takahashi \etal  have studied the magnetoresistive effects of
the F/S/F double tunnel junction caused by competition between
superconductivity and spin accumulation \cite{takahashi1999}.  They
have shown that the F/S/F
double tunnel junction is a magnetoresistive
device to control superconductivity by applying the bias voltage or
current.  The suppression of the superconductivity by the spin
injection in the double tunnel junction was observed by Dong \etal
\cite{dong1997}.  

In this article, we theoretically study the spin-dependent transport 
of the F/S/F SET transistors at zero temperature with a special
attention to the  parity effect.  We assume that the charging energy
$E_{C}$ is larger than $\Delta$ and neglect
the two-electron tunneling \cite{hekking1993} which becomes important in
the opposite situation ($E_{C}<\Delta$).
In such a system, we have the special region called ``plateau region''
where the tunneling current is dominated by the transition 
rate from the odd-state to the even-state \cite{fazio_book}.  
We show that in the plateau region,  spin accumulation is forbidden by
the superconducting gap  even at finite bias voltage.  
However, the information of the injected spin is carried by the excess 
electron and the TMR exists.  The spin relaxation rate of the
excess electron can be estimated from the TMR.  We also show that
the TMR increases as the size of the island decreases.

\begin{figure}
  \includegraphics[width=\columnwidth]{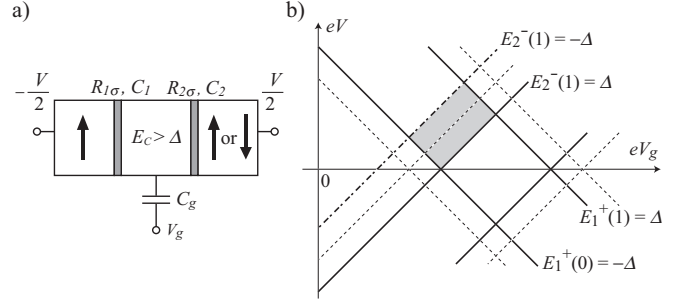}
    \caption{a)Schematic diagram of 
      a F/S/F SET transistor.  The arrows indicate the magnetizations
      of the left and right electrodes.  b)
      The gate and bias voltage diagram of a symmetric
      SET transistors.
      Solid lines indicate the boundaries for the F/S/F SET transistor.  Thin
      dashed lines are the boundaries for the F/N/F SET transistor
      ($\Delta=0$). The dot-dashed line indicates a boundary of the
      plateau region.}
  \label{fig:model}
\end{figure}

We consider the F/S/F SET transistor shown in Fig. \ref{fig:model} (a).
For simplicity we assume that the insulating barriers of
junctions 1(left) and 2(right) are the same; we
subsequently set $C_{1}=C_{2}\equiv C$.  We also assume that the left
and right electrodes are made of the same material with the spin
polarization $P$. 

The Hamiltonian of the SET transistor is given by
\begin{equation}
  H=H_{L}+H_{I}+H_{R}+H_{ch}+H_{TL}+H_{TR},
\end{equation}
where $H_{L}$, $H_{I}$, and $H_{R}$ are the Hamiltonians of the left
electrode, the central island, and the right electrode, respectively.
The Hamiltonian $H_{ch}$ represents the charging energy and  the
tunneling processes are described by 
\begin{equation}
    H_{TL(R)}
    =\sum_{k,q,\sigma} T_{kq}^{L(R)}c_{k\sigma}^{\dag}c_{q\sigma} 
    + {\rm h.c.},
\end{equation}
where the subscript $k$ indicates the wave vector in the left(right)
electrode while $q$ represents that in the 
island and no spin flip is assumed in the tunneling process.

The energy change due to the forward tunneling of an
electron with spin $\sigma$ through the junctions 1 and 2
are respectively given by
\begin{eqnarray}
  E_{1}^{+}(n)
  &=&(1+2n)E_{c} - \frac{C_{g}}{C_{\Sigma}}eV_{g} - \frac{1}{2}eV\\
  E_{2}^{-}(n)
  &=&(1-2n)E_{c} + \frac{C_{g}}{C_{\Sigma}}eV_{g} - \frac{1}{2}eV,
\end{eqnarray}
where $n$ is the number of excess electrons of the initial state
in the island, $C_{\Sigma}\equiv 2C + C_{g}$, and the superscripts $\pm$
implies that the number of excess electrons in the final state
is $n\pm1$. The energy changes for the backward tunneling are given by
$E_{1}^{-}(n) = E_{2}^{-}(n) + eV$ and $E_{2}^{+}(n) = E_{1}^{+}(n) + eV$.

In ordinary tunnel junctions with negligible charging effect,
individual electrons are unstable for the forward tunneling across the
junction, and lower their energy by the bias potential ($-eV/2 <0$).  
In SET transistors, however, there are special regions called CB
region in the gate and bias voltage diagram, where the charge state of
the island with $n$ excess electrons is stable with respect to tunneling.
For the symmetric SET transistor
with a normal conducting island, the CB regions are the rhombuses
determined by  $  E_{1}^{+}(n),  E_{1}^{-}(n),  E_{2}^{+}(n)$, and
$E_{2}^{-}(n) \ge 0$.  The boundaries of CB regions are indicated by
thin dashed lines in Fig. \ref{fig:model} (b).

For the superconducting island, however, it is know that these
boundaries are modified by superconductivity.  In order to obtain
the modified CB regions, we evaluate the tunneling rates following Fazio
and Sch{\"o}n \cite{fazio_book}.
Since the island is in the superconducting state, it is convenient
to rewrite the electron operators in the island
in terms of the quasiparticle operators by using the Bogoliubov
transformations.
Then the transition rates are determined by using the Golden-rule
arguments \cite{fazio_book}.  Let us first consider an electron tunnels
from the left electrode into the island with $n=0$, thereby changing the electron
number from $n=0$ to $n=1$ with the rate
\begin{equation}
  \begin{aligned}
  \Gamma_{1\sigma}^{+}(0)&=\frac{1}{e^{2}R_{1\sigma}}
  \int_{-\infty}^{\infty}dE
  \int_{-\infty}^{\infty}dE^{\prime}
  {\cal D}(E^{\prime})\\
  &\times f(E)\left[1-f(E^{\prime})\right]
  \delta(E^{\prime} - E + E_{1}^{+}(0)),
  \end{aligned}
\end{equation}
where ${\cal D}(E)\equiv |E|/\sqrt{E^{2}-\Delta^{2}}$ is the normalized
BCS density of states and $f(E)$ is the Fermi distribution function.
The tunnel conductance of the junction is defined as
$R_{1\sigma}^{-1}\equiv(4\pi
e^{2}/\hbar)N_{\sigma}^{I}N_{\sigma}^{L}|T|^{2}$, where
the tunnel matrix elements $T_{kq}^{L}$ and $T_{kq}^{R}$
are considered as a constant $T$ and $N_{\sigma}^{I(L)}$ denotes the
density of states of the island in the normal conducting state(left
electrode).
Since we assume the temperature is zero, the Fermi distribution
function can be replaced by the step function and we have
\begin{equation}
  \begin{aligned}
    \Gamma_{1\sigma}^{+}(0)
    &=\left\{
      \begin{array}{cc}
        0 & (E_{1}^{+}(0)\ge -\Delta)\\
        \displaystyle\frac{\sqrt{E_{1}^{+}(0)^{2} 
            - \Delta^{2}}}{e^{2}R_{1\sigma}}  & 
        (E_{1}^{+}(0)< -\Delta)
        \end{array}
        \right.
  \end{aligned}.
\end{equation}
Therefore, the boundary of the CB region for $n=0$ given by
$E_{1}^{+}(0)=0$ is lifted to the solid line determined by
$E_{1}^{+}(0)=-\Delta$ as shown in Fig. \ref{fig:model} (b).
The other three boundaries also shift outward
and the CB region is determined by
$  E_{1}^{+}(0),  E_{1}^{-}(0),  E_{2}^{+}(0)$, and $E_{2}^{-}(0) \ge
-\Delta$.

Next we consider an electron tunnels from the island with $n=1$ to the
right electrode, thereby changing the electron number from $n=1$
to $n=0$ with the rate
\begin{equation}
  \begin{aligned}
  \Gamma_{2}^{-}(1)&=\frac{1}{e^{2}R_{2\sigma}}
  \int_{-\infty}^{\infty}dE
  \int_{-\infty}^{\infty}dE^{\prime}
  {\cal D}(E)\\
  &\times f(E-\delta\mu)\left[1-f(E^{\prime})\right]
  \delta(E^{\prime} - E + E_{2}^{-}(1)),
  \end{aligned}
\end{equation}
where $R_{2\sigma}^{-1}=(4\pi e^{2}/\hbar)N_{\sigma}^{I}N_{\sigma}^{R}|T|^{2}$ and $\delta\mu$ is the shift of the chemical potential fixed by the
constraint of one excess electron charge in the island \cite{fazio_book}.
The shift of the chemical potential is $\delta\mu=\Delta$ and we have
\begin{equation}
  \begin{aligned}
    \Gamma_{2\sigma}^{-}(1)
    &=\left\{
      \begin{array}{l}
        0   \mbox{\hspace{7.7em}} (E_{2}^{-}(1)\ge \Delta)\\
        \displaystyle\frac{d}{e^{2}R_{2\sigma}}
        \mbox{\hspace{5.5em}}
        (|E_{2}^{-}(1)|< \Delta)\\ 
        \displaystyle\frac{d}{e^{2}R_{2\sigma}} +  
        \displaystyle\frac{\sqrt{E_{2}^{-}(1)^{2} 
            - \Delta^{2}}}{e^{2}R_{2\sigma}}  \\
        \mbox{\hspace{8.5em}}(E_{2}^{-}(1)\le -\Delta)
        \end{array}
        \right.
  \end{aligned},
\end{equation}
where $d=1/N_{\sigma}^{I}$ is the average level spacing of the island.
Therefore, the boundary of the CB region for $n=1$ given by
$E_{2}^{-}(1)=0$ is also moved to the solid line determined by
$E_{2}^{-}(1)=\Delta$ as shown in
Fig. \ref{fig:model} (b).  The other three boundaries
also shift inward and the CB region is
determined by 
$  E_{1}^{+}(1),  E_{1}^{-}(1),  E_{2}^{+}(1)$, and $E_{2}^{-}(1) \ge \Delta$.
One can easily show that the CB regions for the other even(odd) states
are spread(squeezed) like that for $n=0(1)$.

Adjacent to the CB region, we have a so-called ``plateau region'' where the 
tunneling current is dominated by the tunneling rate through one
junction which behaves like a bottleneck of the tunneling current.
In the plateau region the tunneling current is carried by the even and
the odd states in the following manner.
An electron with spin $\sigma$ tunnels into the
island from the left electrode.  While staying in the
island, the spin of the electron relaxes due to the 
spin orbit interaction, surface scattering,
and/or the hyperfine contact interaction \cite{yafet1983}.  
After a certain time period determined by the tunneling rate
$\Gamma_{2\sigma}^{-}(1)$, the electron tunnels out of the
island.  Therefore, no spin accumulation occurs even at the finite
bias voltage.  Outside the plateau region, continuous quasiparticle
states contribute to the tunneling current and the spin accumulation
can exist.

We consider the tunneling current and TMR in the
plateau region, for example, indicated by shade in
Fig. \ref{fig:model} (b), which is determined by 
$E_{1}^{+}(0) <-\Delta, E_{1}^{+}(1) > \Delta$, and $|E_{2}^{-}(1)| <
\Delta $.
In this region, the following three states are available: $|0\rangle$
three is no excess electron in the island, $|\uparrow\rangle$ there is one
up-spin excess electron in the island, $|\downarrow\rangle$ there is one
down-spin excess electron in the island.  
The transition rate from $|0\rangle$ to $|\sigma\rangle$ is given by
\begin{equation}
  \Gamma_{\sigma}^{+}\equiv\Gamma_{1\sigma}^{+}(0) = \frac{1}{e^{2}
  R_{1\sigma}}\sqrt{{E_{1}^{+}(0)}^{2} - \Delta^{2}},
\end{equation}
and  the transition rate from  $|\sigma\rangle$ to $|0\rangle$ is
\begin{equation}
  \Gamma_{\sigma}^{-}\equiv\Gamma_{2\sigma}^{-}(1) = \frac{d}{e^{2}
  R_{2\sigma}}.
  \label{eq:gamma2}
\end{equation}
We also introduce the spin relaxation rate $\eta$ which corresponds to the
transition rate between $|\uparrow\rangle$ and $|\downarrow\rangle$.

In order to obtain the tunneling current, we construct the master
equation for the probabilities of states $p_{0}$, $p_{\uparrow}$, and
$p_{\downarrow}$, which is given by
\begin{eqnarray}
 &&\dot{p}_{0}
 =\Gamma_{\uparrow}^{-}p_{\uparrow}+\Gamma_{\downarrow}^{-}p_{\downarrow}
 -\left(\Gamma_{\uparrow}^{+}+\Gamma_{\downarrow}^{+}\right)p_{0}\\
&& \dot{p}_{\sigma}
 =\eta\left(p_{\bar{\sigma}} - p_{\sigma}\right) + \Gamma_{\sigma}^{+}p_{0}
 -\Gamma_{\sigma}^{-}p_{\sigma}
\end{eqnarray}
with the normalization condition $p_{0}+p_{\uparrow}+p_{\downarrow}=1$.
We calculate the stationary probabilities by requiring
$\dot{p}_{0}=\dot{p}_{\uparrow}=\dot{p}_{\downarrow}=0$.
The solutions are
\begin{eqnarray}
  p_{0}&=&\frac{\eta(\Gamma_{\uparrow}^{-}+\Gamma_{\downarrow}^{-})
  + \Gamma_{\uparrow}^{+}\Gamma_{\downarrow}^{+}}{\gamma}\\
  p_{\sigma}&=&\frac{\eta\left(\Gamma_{\sigma}^{+}+\Gamma_{\bar{\sigma}}^{+}\right)
  + \Gamma_{\sigma}^{+}\Gamma_{\bar{\sigma}}^{-}}{\gamma},
\end{eqnarray}
where
$\gamma=\sum_{\sigma=\uparrow\downarrow}\eta\left(2\Gamma_{\sigma}^{+} + 
\Gamma_{\sigma}^{-}\right) + \Gamma_{\sigma}^{+}\Gamma_{\bar{\sigma}}^{-}
 + \frac{1}{2}\Gamma_{\sigma}^{-}\Gamma_{\bar{\sigma}}^{-}$ and the subscript $\bar{\sigma}$ represents the spin direction opposite
to $\sigma$.  The similar result has been obtained for the single
discrete level system\cite{rudzinski2001}.
Knowing the stationary probabilities we can calculate the
tunneling current through the left junction
\begin{equation}
  I=-e p_{0}\left(\Gamma_{\uparrow}^{+} + \Gamma_{\downarrow}^{+}\right).
  \label{eq:current}
\end{equation}

The tunneling current of the F/S/F SET transistor given by
Eq. (\ref{eq:current}) depends strongly on whether the magnetizations in 
ferromagnetic electrodes are parallel or antiparallel.  
For the ferromagnetic(F)-alignment, where the
magnetizations are parallel, the junction resistances can be expressed
as $R_{1\uparrow} = R_{2\uparrow}= R_{M}$ and
$R_{1\downarrow} = R_{2\downarrow}= R_{m}$.
Here $R_{M}$ is the junction resistance for electrons in the majority
spin band and $R_{m}=(1-P)/(1+P)R_{M}$ is for those in the minority spin band.
We write the transition rates by using the subscripts $M$ and $m$ as
$  \Gamma_{\uparrow}^{\pm} = \Gamma_{M}^{\pm}$ and $ \Gamma_{\downarrow}^{\pm}
 =\Gamma_{m}^{\pm}$.
Introducing $\alpha\equiv R_{m}/R_{M}$, $\beta\equiv
\Gamma_{M}^{-}/\Gamma_{M}^{+}$,$\zeta\equiv \eta/\Gamma_{M}^{-}$, the
probability of the state $|0\rangle$
for the F-alignment is expressed by
\begin{equation}
  p_{0}^{F} =
  \frac{\zeta \beta (\alpha + 1) + \alpha \beta}{\zeta (\alpha +
    1)(\beta + 2) + \alpha (\beta + 2)}=\frac{\beta}{\beta+2}.
  \label{eq:pf}
\end{equation}
The probability $p_{0}^{F}$ is independent of the spin relaxation rate.
For the antiferromagnetic(A)-alignment, where the magnetizations are
antiparallel, the probability of the even state is
\begin{equation}
  p_{0}^{A} =
  \frac{\zeta \beta (\alpha + 1) + \alpha \beta}{\zeta (\alpha +
  1)(\beta + 2) + \alpha^{2} + \alpha\beta + 1}.
  \label{eq:pa}
\end{equation}

\begin{figure}
    \includegraphics[width=\columnwidth]{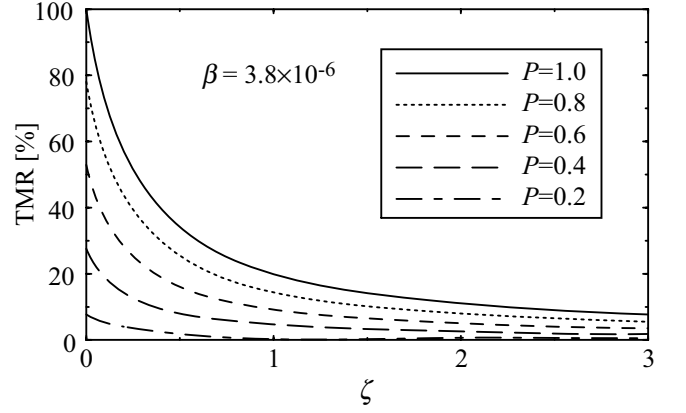}
    \caption{TMR of the F/S/F SET with $\beta=3.8\times 10^{-6}$ is
      plotted against the normalized spin relaxation rate $\zeta$.
      From top to bottom: $P=1.0, 0.8, 0.6, 0.4$, and 0.2.}
  \label{fig:zeta-dep}
\end{figure}

The TMR is calculated by  the formula
$TMR=1-I_{A}/I_{F}$. From Eqs. (\ref{eq:current})-
(\ref{eq:pa}), we have 
\begin{eqnarray}
  TMR=\frac{(\alpha-1)^{2} }{
    \zeta(\alpha +1)(\beta +2) + \alpha^{2} + \alpha\beta + 1}.
  \label{eq:tmr}
\end{eqnarray}
The value $\beta$ is much smaller than unity in almost all area of the plateau
region since $d \ll E_{C}$.  If we assume that $\Delta$=0.18 meV (Aluminum),
$E_{C}$=8.0meV and $d=3.0\times 10^{-5}$ meV \cite{ralph_book}, then we
have $\beta=3.8\times 10^{-6}$ at the center of the plateau region. 
In Figure \ref{fig:zeta-dep}, the TMR for $\beta=3.8\times
10^{-6}$ is plotted against the normalized spin
relaxation rate
$\zeta$.  The TMR is monotonically decreasing function of $\zeta$ and 
the spin relaxation rate of an excess electron $\eta$  can be
estimated from the TMR by fitting the experimental data.  If there is
no spin relaxation process in the island, $\zeta=0$, the TMR is
approximately given by $TMR \simeq 2P^{2}/(1+P^{2})$.

The fact that the TMR depends on the spin relaxation rate $\eta$ via
its normalized value $\zeta$ means that how much the spin information is
transmitted is determined by the competition between the spin relaxation
rate and the tunneling rate through the right junction. 
In the plateau region the inverse of the tunneling rate
$\Gamma_{M(m)}^{-}$ describes how long the excess electron with
majority(minority) spin stays in the island.  
Therefore, the normalized spin relaxation rate $\zeta$
represents the probability that the electron with the majority spin
tunnels out of the island holding its spin direction and the TMR is a
function of $\zeta$.

\begin{figure}
  \includegraphics[width=\columnwidth]{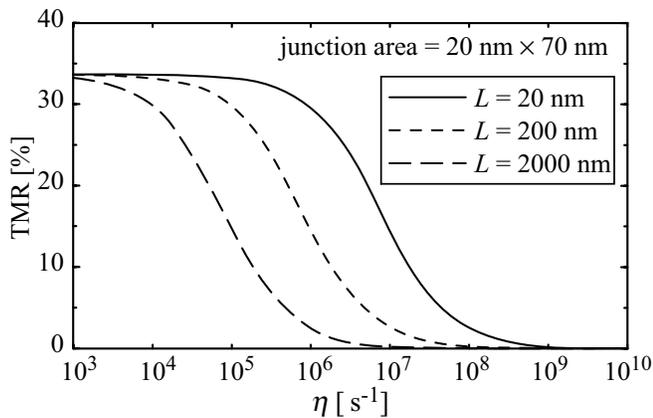}
    \caption{TMR of the F/S/F SET with $P=0.45$, which is the typical
      value for Co \protect\cite{moodera1999},  is
      plotted against the spin relaxation rate $\eta$.
      From top to bottom: the length of the island is $L$= 20, 200,
      and 2000 nm.  The area of the junction is fixed at 20nm $\times$ 70
      nm\protect \cite{ralph_book} and the junction resistance $R_{M}$
      = 1 M$\Omega$ for $L$=2000 nm.  We assume that $\Delta$=0.18 meV
      (Aluminum), $E_{C}$=8.0meV and $d=3.0\times 10^{-5}$ meV and the
      working point is set to the center of the plateau region.}
  \label{fig:L-dep}
\end{figure}

In the WKB approximation, the value $|T|^{2}$ is inversely proportional
to the length of the island $L$ \cite{bardeen1961,harrison1961}.  If
the junction parameters other than $L$ are kept fixed, the density of
states $N_{\sigma}^{I}(=1/d)$ is proportional to $L$.
In this situation, the tunneling rate through the left junction 
does not depend on $L$ since the size
dependences of $|T|^{2}$ and $N_{\sigma}^{I}$ in $R_{\sigma}^{1}$
cancel out. On the contrary, the tunneling rate through the right junction is
inversely proportional to $L$.   The normalized spin relaxation rate
$\zeta$ decreases and therefore the TMR increases with decreasing $L$. 
In Fig. \ref{fig:L-dep} the TMR of the
F/S/F SET transistor with $P=0.45$, which is the typical value for Co \cite{moodera1999}, is plotted against the spin
relaxation rate $\eta$ for various values of $L$.
We assume that the area of the junction is fixed at 20 nm $\times$ 70 nm
and $L$=2000 nm, 200, and  20 nm corresponding to the Aluminum island
with the average level spacing $d = 3.0 \times 10^{-5}$, $3.0 \times 10^{-4}$,
and $3.0 \times 10^{-3}$ meV, respectively \cite{ralph_book}.  
For the spin relaxation rate $\eta = 10^7$ s$^{-1}$,
which is of the same order as that  caused by the hyperfine contact
interaction \cite{yafet1983}, the TMR is 0.26, 2.4, and 15 \% for $L$ =
2000, 200, and  20 nm, respectively.

Very recently, Chen \etal reported the experimental
evidence for suppression of superconductivity by spin imbalance in
Co/Al/Co SET transistors\cite{chen2002}.  Although they observed
the spin imbalance outside the plateau region, their experimental results
show that the effect we proposed is relevant to current experiments.

In conclusion, we theoretically study the TMR
of F/S/F SET transistors with $E_{C} > \Delta$. We show that in the
plateau region, there is no spin accumulation even at the finite bias
voltage.  However, the information of the injected spin is carried by the
excess electron and the TMR exists.  The spin relaxation rate of the
excess electron can be estimated from the TMR.  We also found that the
TMR increases with decreasing the size of the superconducting island.

We would like to thank S. Maekawa and S. Takahashi for
valuable discussions.  
This work was supported by a Grant-in-Aid for Scientific
Research (C), No. 14540321 from MEXT. H.I. was supported by MEXT,
Grantin-Aid for Encouragement of Young Scientists, No.
13740197.

\end{document}